# Optimal Relay Location in Diffusion Based Molecular Communications

**Ghazaleh Ardeshiri**
School of Electrical and Computer Engineering
Shiraz University
Shiraz, Iran
gh.ardeshiri@shirazu.ac.ir

**Ali Jamshidi**
School of Electrical and Computer Engineering
Shiraz University
Shiraz, Iran
jamshidi@shirazu.ac.ir

**Alireza Keshavarz-Haddad**
School of Electrical and Computer Engineering
Shiraz University
Shiraz, Iran
keshavarz@shirazu.ac.ir

**Abstract**: *Molecular Communications via Diffusion (MCvD) is a promising paradigm which enables nano-machines to communicate with each other. However, the reliability of existing systems degrades rapidly as the distance between the transmitters and the receivers grows. To solve this issue, relaying schemes must be implemented in practice. In this paper, we study two relaying schemes: In the first case, the relay node decodes the incoming signal symbol and forwards it to the receiver using a different type of molecule. Then, the receiver detects the information bits by only considering the molecules from the relay node. In the second case, the receiver considers both the types of molecules sent from the transmitter and the relay node. For these two scenarios, the optimal location of the relay node are obtained. We assume Quadruple Concentration Shift Keying (QCSK) modulation in which the signal is encoded into the four level concentrations of molecules emitted by the nano-machines. Our simulation results indicate that adding a relay improves the performance by 10dB and 15dB in the first and the second schemes, respectively.*

## I. INTRODUCTION

Nanotechnology enables the development of devices in a scale ranging from one to a few hundred nanometers. At nano scale, the most basic functional unit is a nano-machine, which is able to perform simple tasks such as computing, data storing, sensing or actuation [1]. One of the important areas of nano-networks is the biomedical domain, which includes health monitoring, tissue engineering, and targeted drug delivery [2]. Other application domains include industrial applications, such as new materials and quality control of products, and environmental applications, such as biodegradation and air pollution control [3], [4].

In molecular communication, signal is encoded and decoded by molecules rather than electromagnetic waves [5]. Molecular Communication via Diffusion (MCvD) is the most promising approach for the communication between nano-machines [6]. In MCvD, the molecules that are released by the transmitter nano-machine in a fluid environment randomly walk in all directions without any further infrastructure and some of them may reach the receiver nano-machine [2].

One of the main drawbacks of MCvD is its limited range of communication, since the propagation time increases and the number of received molecules decreases with growing the distance. This makes communication over larger distances challenging [7]. One approach in conventional wireless communications that can be adapted for MCvD is the use of intermediate transceivers acting as relays to aid the communication with distant receivers. Such relays can potentially improve the reliability and performance of the communication [2].

In this paper, we investigate two different relaying schemes. The relay node is placed between the transmitter and the receiver decodes the incoming signal and forwards it to the receiver using a different type of molecule than the one used by the transmitter. In the first case, the receiver uses only molecules which come from the relay node and ignores molecules from the transmitter. In the second case, the receiver considers both molecules received from the transmitter and the relay node.

For each scheme optimal location of the relay node is obtained. Interestingly, the optimal location for the first scheme is the middle point between the transmitter and the receiver, while for the second scheme the relay should be placed near to the transmitter.

Further, we assume that Quadruple Concentration Shift Keying (QCSK) modulation for the communications among the transmitter, the relay, and the receiver. The optimal concentrations and thresholds for QCSK modulation are calculated by Monte Carlo simulations. Our simulation results indicate that for the first case the Symbol Error Rate (SER) is improved by 10dB and in the second case up to 15dB improvement is achieved.

The remainder of the paper is organized as follows. In section II, the MCvD system model is introduced, and QCSK is described. In section III, we introduce our relay schemes. Section IV, provides simulation results of the optimal concentrations and thresholds for QCSK modulation and the relay node. Finally, conclusion is drawn in section V.

II. SYSTEM MODEL

*A. Diffusion and the First Hitting Time*

The molecules are the information particles in molecular scale. In this scale, the movement of particles inside a fluid is modeled by Brownian motion or diffusion process. If we focus on the diffusion process of a particle starting from origin, then the concentration at radius *r* and time *t* is given by the following formula [8]:

$$C(r,t) = \frac{1}{(4\pi Dt)^{n/2}} e^{-r^2/4Dt} \quad (1)$$

where *n* and *D* are the dimension of the environment and the diffusion coefficient, respectively [9]. In nature, whenever a messenger molecule hits the body of the receiver, the molecule is received and removed from the environment; therefore, the hitting molecule cannot move further and constitutes the signal just once. This process is referred to as first passage or the hitting process. What we are concerned with the probability that a diffusing particle first reaches a specified site at a specified time [10].

The first generalized model for probability distribution function of first hitting process in 1-D environment was derived as [11]

$$f_{hit}^{1D}(t) = \frac{d}{\sqrt{4\pi Dt^3}} e^{-d^2/4Dt} \quad (2)$$

where *d* corresponds to the distance. Therefore, the probability of hitting an absorbing receiver until time *t*, can be obtained from integrating of (2) as follows [12]

$$F_{hit}^{1D}(t) = \text{erfc}\left(\frac{d}{\sqrt{4Dt}}\right) \quad (3)$$

Similarly, fraction of hitting molecules to a perfectly absorbing spherical receiver in a 3-D environment is derived in [13]. Hitting rate of molecules to a spherical receiver in a 3-D environment is formulated as

$$f_{hit}^{3D}(t) = \frac{r_r}{d+r_r} \frac{d}{\sqrt{4\pi Dt^3}} e^{-d^2/4Dt} \quad (4)$$

where $r_r$ denotes the radius of spherical receiver. One can obtain the fraction of hitting molecules until time *t* by integrating $f_{hit}^{3D}(t)$ in (4) with respect to time, which yields similar results with the 1-D case.

Note that there is a positive probability of no hitting to the absorbing boundary for a diffusing particle in a 3-D environment when time goes to infinity. The survival probability depends on the radius of the receiver and the distance between the transmitter and the receiver [13].

*B. Concentration Shift Keying (CSK) Modulation*

The concentration of the received molecules is used as the amplitude of the signal. In order to represent different symbols, the transmitter releases different number of molecules for each symbol. For example, for "0" the transmitter releases $N_0$ molecules whereas for "1" $N_1$ molecules will be released [14]. Then, the receiver detects the intended symbol as "1" if the number of molecules arriving at the receiver during a time slot exceeds a threshold $\tau$ (assuming $N_1 > N_0$). Otherwise, the symbol is detected as "0".

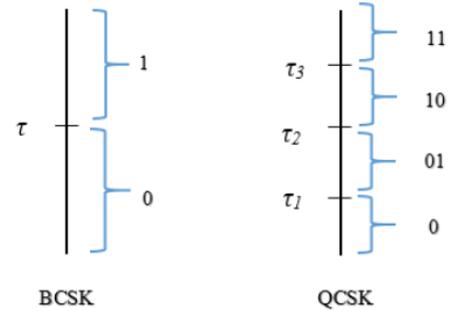

Figure 1. BCSK and QCSK modulations' thresholds

CSK is analogous to Amplitude Shift Keying (ASK) in classical communication. Instead of using two values, e.g., $N_0$ and $N_1$, and a single threshold, the symbol can be tailored to represent *b* bits by using $2^b$ different values with $2^b -1$ threshold levels. CSK can be implemented in practice as BCSK (Binary CSK) or QCSK (Quadruple CSK), depending on the bits per symbol rate.

- If *b = 1*, CSK is called Binary CSK (BCSK)
- If *b = 2*, CSK is called Quadruple CSK (QCSK).

Extension BCSK to higher order, e.g. QCSK, is not as easy as conventional communication. We can consider four different level of concentrations for the symbols, i.e. $N_0, N_1, N_2, N_3$. Notice that, in general the optimal thresholds ($\tau_1$, $\tau_2$ and $\tau_3$ in Figure 1) are not midpoint of the molecule quantity per symbol, which is consistent with the theoretical analysis.

In this work, for the first time the optimal concentrations and thresholds for this modulation are investigated. For simplifying our simulations, we assume that for QCSK modulation the transmitter does not send any molecules for symbol "0", sends *N* molecules for symbol "1", 2*N* molecules for symbol "2", 3*N* molecules for symbol "3". Next, we find the optimal concentration *N* for a given baud-rate via simulation.

*C. ISI Model*

Arising from the probabilistic dynamics of Brownian motion, the signal molecules move randomly and do not necessarily reach the receiver, moreover the arrival of molecules spreads to a very long duration. Instead, every molecule has a probability of hitting the receiver in a predetermined time duration $t_s$. The received molecules contain ISI due to surplus molecules from the previous symbols and affects the decoding process severely [8].

### D. Noise Model

Channel noise is mainly caused by other nano-machines, apart from the possible molecular reaction and background molecules. Receiving and counting undesired molecules in demodulation, could be regarded as noise. Positive noise means receiving redundant molecules from other nano-devices, negative noise means some molecules are received by other nano-devices [15]. In most papers, it is assumed that the noise is Additive White Gaussian Noise (AWGN) and expressed as,

$$N_{noise}(n) \sim Normal(0, \sigma^2) \quad (5)$$

Notice that, in this system model, the noise takes discrete values, however, for simplification, its distribution function is approximated as (5). Here, the noise power is defined as the variance of the normal distribution.

### III. RELAY SCHEMES

In this section, we study the effect of relaying in MCvD. We assume that the transmitter is a point source and placed at location *(0, 0, 0)* and the receiver and the relay nodes spherical in shape with fixed volumes and radii in 3-D space. Also, they are passive observers such that molecules can diffuse through them [7]. The receiver is placed at the distance of $6\mu m$ from the transmitter. The relay node is placed between the transmitter and the receiver.

We assume that there are two distinct types of molecules, type I and type II, the transmitter releases type I molecules. The relay node can detect type I molecules and emits type II molecules. The receiver can detect both types. Our goal is to minimize SER by optimally choosing the location of the relay node.

We use the SNR definition presented in [14]; the ratio between the average received power and the average noise power. The transmission power is defined as the number of molecules sent by the transmitter. Detection process in relay node is based on the optimal thresholds explained before and we use the optimal concentration for each distance ($d_{12}$, $d_{13}$ and $d_{32}$ in Figure 2). The following cases consider two detection processes for the receiver node.

*First relaying scheme:* In this case, the transmitter releases defined concentration of type I molecules for each symbol. The relay node can diagnose type I molecules and by comparing the number of received molecules in each time slot with the thresholds, it can detect symbols. The Optimal thresholds based on the value of $d_{12}$ are determined for this case. The relay node releases concentration of type II molecules for each symbols. Detection process in the receiver carries out just using type II molecules from the relay node. The receiver uses optimal thresholds based on the value of $d_{32}$.

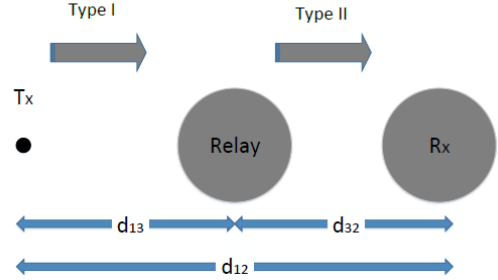

Figure 2. The Relaying Scheme

*Second relaying scheme:* In the second case, the receiver does not ignore type I molecules received from the transmitter, and uses Maximum A Posteriori (MAP) detection based on the two received types. In this paper, performance of the relaying system is investigated by using this kind of detection. By conducting simulations, first, we obtain decision boundaries for different location of the relay node. Then, we compare SERs for the different relay locations to specify the optimal location of the relay node.

### IV. PERFORMANCE EVALUATION

In this section, we study the performance of our proposed relaying schemes. First, we obtain the parameters of QCSK modulation optimally. Next, we present simulation results for the relaying schemes. In this section, all simulation results are provided with parameters given in Table I.

### A. Optimal thresholds

To minimize symbol error rate, optimal thresholds for QCSK modulation are essential. We obtain these optimal thresholds for different distances by conducting some simulations. Notice that, we assume the transmitted symbols have the same probability. The probability density function (PDF) of the received molecules can be obtained via simulation. These four PDFs have three incidence points (intersection of nearby PDFs) which can be used as an approximation for the optimal threshold values. For example, Figure 3 depicts the PDFs corresponding to various symbols for $d = 3\mu m$ and $N = 150$ over a noiseless channel. The incidence point of PDFs corresponding to symbols "0" and "1", is *(x, y) = (107.8, 0.003)*. The value of x-axis shows approximated optimum threshold $\tau_1$.

If the hitting molecules by the receiver is less than 108, symbol "0" is decoded.

The 3rd Iranian Conference on Communications Engineering (ICCE 2017)
Shahid Rajaee University, Tehran, Iran, 23-24 February 2017

TABLE I. SIMULATION PARAMETERS

| Parameter | Simulation Parameters |
|---|---|
| $D$ | 100μm²/s |
| $R_{relay}$, $R_{receiver}$ (radii) | 4μm |
| Symbol duration | 0.15s |
| Sampling duration | 0.15s |
| Number of symbols | 50000 |
| Distance | 6μm |

TABLE II. THRESHOLDS

| Distance | Thresholds | | |
|---|---|---|---|
| | $\tau_1$ | $\tau_2$ | $\tau_3$ |
| 1μm | 80 | 213 | 345 |
| 2μm | 96 | 210 | 318 |
| 3μm | 108 | 198 | 287 |
| 4μm | 116 | 187 | 285 |
| 5μm | 122 | 180 | 235 |
| 6μm | 127 | 177 | 214 |

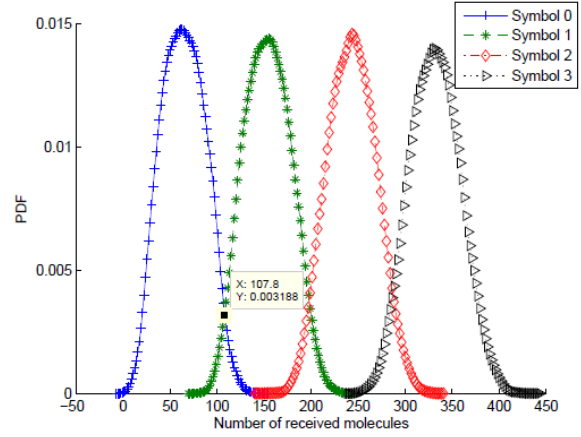

Figure 3. Thresholds for distance $3\mu m$ and $N$=150

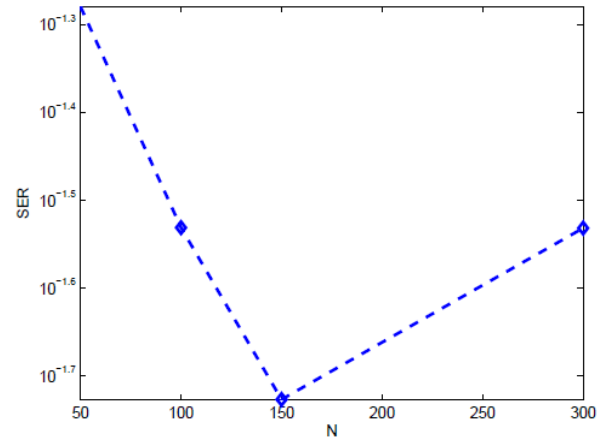

Figure 4. SER of various concentrations for distance $3\mu m$ ($SNR = \infty$)

Table II includes the threshold values when the distance between the transmitter and the receiver varies from $1\mu m$ to $6\mu m$.

Note that it is not necessarily in the midpoint of the released concentration per symbol. This shows that as distance increases, the value of $\tau_1$ increases and the values of $\tau_2$ and $\tau_3$ decrease. For distance $1\mu m$, $\tau_1$ = 80, $\tau_2$ = 213 and $\tau_3$ = 345. For distance $6\mu m$, $\tau_1$ = 127, $\tau_2$ = 107 and $\tau_3$ = 214. As you can see, the optimal thresholds for distance $6\mu m$ are closer to each other, so, detection process for this distance is harder and error probability is higher. In the next subsection, we obtain for $N = 50, 100, 150$, and $300$, to compute the optimal concentration values.

B. *Optimal Concentration*

Concentration level affects error probability directly. Reducing concentrations, reduces the overall effect of ISI, and increases the probability of miss detection and increasing concentrations enhances the ISI which in turn increases the probability of false alarm, so, some optimal concentrations must be utilized. We conducting simulations to compute these concentrations for different distances. Figure 4 depicts SER for $N = 50, 100, 150, 300$ and $d = 5\mu m$ over noiseless channel when the threshold values are chosen optimally. Moreover, Figure 5 depicts SER when the channel is noisy and SNR varies from -5dB to 40dB. Both results show that that $N = 150$ has the minimum SER among all cases.

C. *Relaying Scheme*

Here we assume that the concentration parameter is N =150 and the threshold values are chosen optimally based on distances. The distance between the transmitter and the receiver is $6\mu m$. We examine the performance of the system by placing the relaying node at distances $2\mu m$, $3\mu m$ and $4\mu m$ from the transmitter. Figure 6 depicts SER for the first relaying scheme. Optimal concentrations are utilized by the transmitter and the relay node. The receiver and the relay node use optimal thresholds in detection processes. Noisy channel is considered and SNR varies from -10dB to 15dB. As Figure 6 depicts, SER reaches its minimum value for distance $3\mu m$. For the second relaying

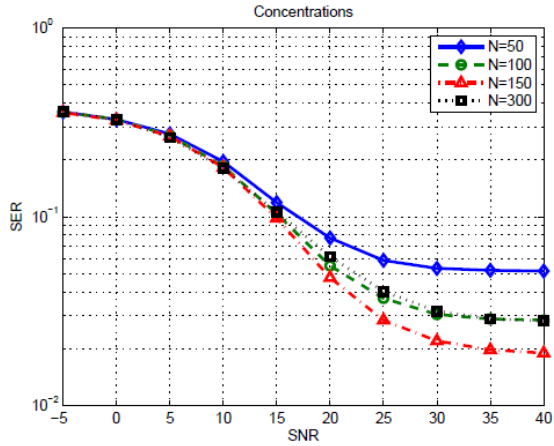

Figure 5. SER of various concentrations vs. SNR for distance $3\mu m$

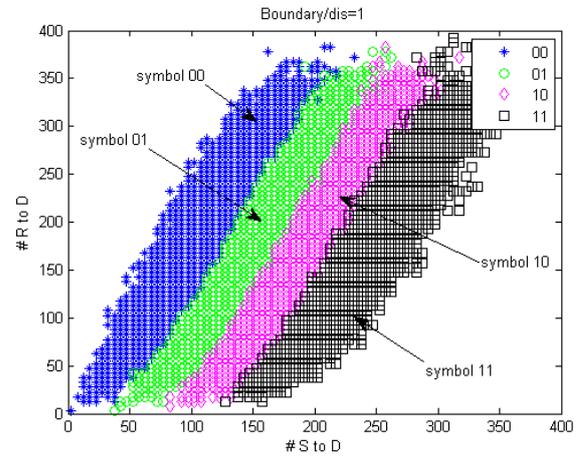

Figure 7. Decision's boundaries at distance $1\mu m$, $N = 150$

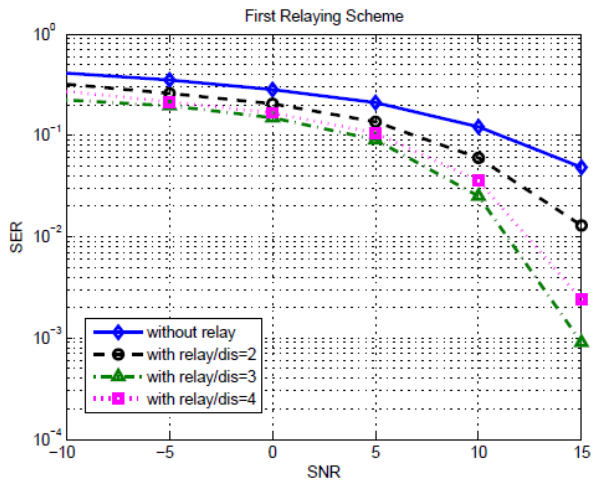

Figure 6. SER of first relaying scheme

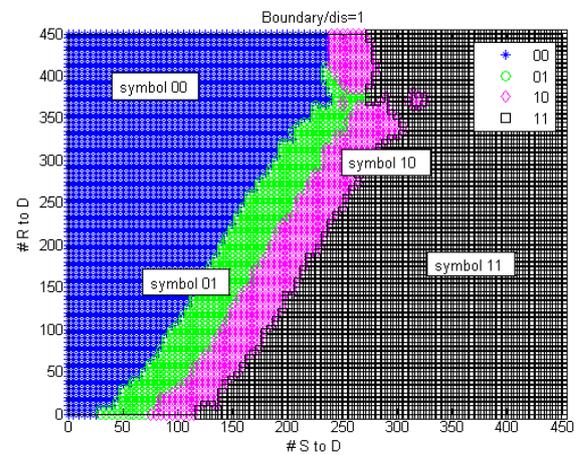

Figure 8. Expanded Decision's boundaries at distance $1\mu m$, $N = 150$

scheme, we need to specify decision's boundaries (regions) for the receiver.

In Figure7 x-axis and y-axis are the number of type I and type II molecules which are received by the receiver, when the relay node is placed at distanced $1\mu m$ from the transmitter and the channel is noiseless. We use four markers to clarify the boundaries and regions for the symbols. In Figure 8, we expand regions of each symbol which are obtained from the last simulation, so, we can use it in noisy channel as well.

Next, we conduct simulations when the relay node is placed at distance 2, 3, 4, 5 $\mu m$ from the transmitter and obtain the optimal boundaries. Then, we compute SER for each case. Figure 9 depicts SER when the relay node is located at different distances and SNR varies from -10dB to 15dB.

This figure indicates that when the relay node is in distance $1\mu m$ from the transmitter, the performance of this relaying scheme is higher than other cases.

In Figure10 we compared different relay schemes with our proposed scheme which is based on MAP detector. In [7], amplify and forward relaying was presented. In this paper, the relay node was placed in the middle between the transmitter and the receiver node and the amplification factor was set as K=50.

Decode and forward relaying has been presented as Multi-Molecule multi-Hop Network (MM-MH) in [2]. MM-MH is the same as our first relaying scheme. Based on [2] and our simulations result in Figure6, we locate the relay node in the middle of distance between the transmitter and the receiver, $3\mu m$ from the transmitter, in order to obtain the best performance for the relay node.

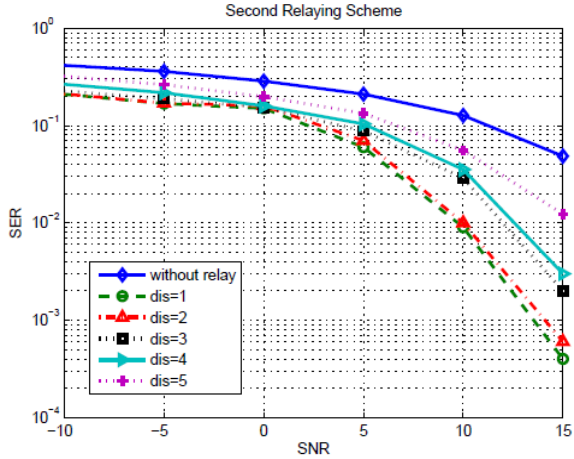

Figure 9. SER of the second relaying scheme

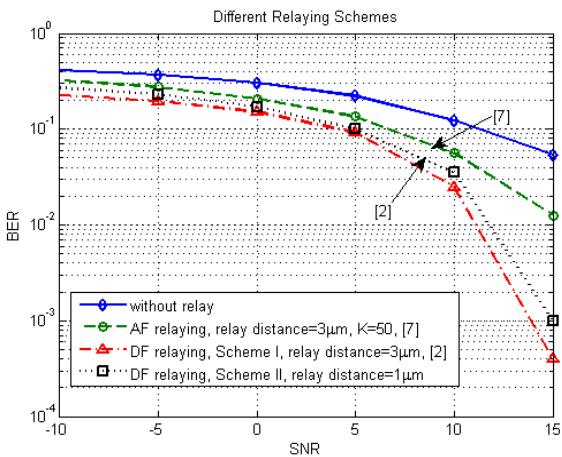

Figure 10. Comparing different relaying schemes

## V. CONCLUSION

In this paper, we considered a three-node network where a nano-relay is deployed between a nano-transmitter and a nano-receiver. We studied two schemes of relaying. In both schemes, simulation results showed that the quality of communication can be significantly improved after deploying relay node. Optimal thresholds and concentration were also calculated to mitigate the effect of ISI. In the first relaying scheme, our simulation results showed that when relay node is in the midpoint of the transmitter and the receiver, SER is minimum. In the second relaying scheme, simulations indicated that the optimal location of the relay node is near to the transmitter.

In this situation, the results indicated that the probability of correct decoding by the relay node is substantially improved. Simulation results showed that our proposed relaying scheme has better performance than existing schemes.